\def\Title{Heisenberg's original derivation of the uncertainty principle
and its universally valid reformulations}
\def\Author{Masanao Ozawa}
\def\Affiliation{Graduate School of Information Science,
Nagoya University, Chikusa-ku, Nagoya, 464-8601, Japan}

\def\Abstract{
Heisenberg's uncertainty principle was originally posed for the limit of the accuracy
of simultaneous measurement of non-commuting observables as stating that
canonically conjugate observables can be measured simultaneously only with 
the constraint that the product of their mean errors should be no less than a limit set 
by Planck's constant. 
However, Heisenberg with the subsequent completion by Kennard has long been credited 
only with a constraint for state preparation represented by the product of the standard deviations.
Here, we  show that Heisenberg actually proved the constraint for the accuracy of 
simultaneous measurement
but assuming an obsolete postulate for quantum mechanics.
This assumption, known as the repeatability hypothesis,
formulated explicitly by von Neumann and Schr\"{o}dinger, was broadly accepted
until the 1970s, but abandoned in the 1980s, when completely 
general quantum measurement theory was established. 
We also survey the author's recent proposal for a universally valid reformulation 
of Heisenberg's uncertainty principle under the most general assumption 
on quantum measurement.
}
\def\Keywords{quantum measurement, uncertainty principle, simultaneous measurement,
repeatability hypothesis, instruments, root mean square error}

\documentclass[prb,twocolumn,showkeys]{revtex4-1}
\usepackage{epsf}
\usepackage{mathptmx}     
\usepackage{amsmath,amssymb,graphicx,color,bm}
\usepackage{amsthm}
\usepackage{enumerate}

\newcommand{\bA}{{\bf A}}
  \newcommand{\beq}{\begin{equation}}
  \newcommand{\beqa}{\begin{eqnarray}}
  \newcommand{\beqas}{\begin{eqnarray*}}
  \newcommand{\beql}[1]{\begin{equation}\label{eq:#1}}
  \newcommand{\bM}{{\bf M}}
\newcommand{\bP}{{\bf P}}
\newcommand{\R}{{\mathbb R}}
\newcommand{\bS}{{\bf S}}
\newcommand{\bx}{{\bf x}}
  
\newcommand{\cB}{{\mathcal B}}
  \newcommand{\cC}{{\mathcal C}}
\newcommand{\cH}{{\mathcal H}}
\newcommand{\cI}{{\mathcal I}}
\newcommand{\cK}{{\mathcal K}}
\newcommand{\cS}{{\mathcal S}}
\newcommand{\da}{\dagger}
 \newcommand{\de}{\delta}
\newcommand{\De}{\Delta}
  \newcommand{\eeq}{\end{equation}}
  \newcommand{\eeqa}{\end{eqnarray}}
  \newcommand{\eeqas}{\end{eqnarray*}}
  \newcommand{\eq}[1]{(\ref{eq:#1})}
  \newcommand{\Eq}[1]{Eq.~(\ref{eq:#1})}
 \newcommand{\et}{\eta}
\newcommand{\Ga}{\Gamma}
 \newcommand{\hp}{\hat{p}}
  \newcommand{\hq}{\hat{q}}
\newcommand{\hy}{\hat{y}}
\newcommand{\la}{\lambda}
\newcommand{\mb}{\mbox}
 \newcommand{\nn}{\nonumber}
 \newcommand{\ran}{\mbox{\rm ran}}
\newcommand{\rh}{\rho}
\newcommand{\si}{\sigma}
\newcommand{\ta}{\tau}
\newcommand{\tc}{\tau c}
\newcommand{\Tr}{\mbox{\rm Tr}}
 \newcommand{\ep}{\varepsilon}

\newtheorem{Theorem}{Theorem}
\newcommand{\hx}{\hat{x}}
\newcommand{\ps}{\psi}
\newtheorem{Axiom}{Axiom}
\newcommand{\av}{\bracket}
\newcommand{\bra}[1]{\langle#1|}
\newcommand{\bracket}[1]{\langle#1\rangle}
\newcommand{\Det}{\De t}
\newcommand{\ie}{{\it i.e.}}
\newcommand{\ket}[1]{|#1\rangle}
\newcommand{\ketbra}[1]{\ket{#1}\bra{#1}}
\newcommand{\px}{\hp_x}
\newcommand{\py}{\hp_y}
\newcommand{\x}{\hx}
\newcommand{\y}{\hy}
\newcommand{\q}{\hat{q}}
\newcommand{\p}{\hat{p}}

\theoremstyle{axiom}

\theoremstyle{postulate}

\renewcommand{\P}{\hat{p}}
\newcommand{\Q}{\hat{q}}
  \newcommand{\ve}{\varepsilon}

 \newcommand{\cL}{{\cal L}}

\newcommand{\id}{{\rm id}}
\newcommand{\sd}[1]{\sigma{(#1)}}

\begin{document}
\title{\Title}
\author{\Author}
\affiliation{\Affiliation}
\email{ozawa@is.nagoya-u.ac.jp}
\begin{abstract}
\Abstract
\end{abstract}
\maketitle

\section{Introduction}
The uncertainty principle proposed by Heisenberg \cite{Hei27E} in 1927
revealed that we cannot determine both position and momentum of a particle
simultaneously in microscopic scale as stating
``the more precisely the position is determined, the less precisely the momentum is 
known, and conversely'' \cite[p.~64]{Hei27E}, 
and had overturned the deterministic world view based on the Newtonian mechanics.
By the famous $\gamma$ ray microscope thought experiment Heisenberg \cite{Hei27E} 
derived the relation
\beql{Hei27E}
\ep(\Q)\ep(\P)\sim h
\eeq
for $\ep(\Q)$, the ``mean error'' of the position measurement,  and $\ep(\P)$,
thereby caused ``discontinuous change'' of the momentum, 
or more generally the mean error of the simultaneous momentum measurement, 
where $h$ is Planck's constant:
\begin{quote}
Let $\ep(\Q)$ [originally, $q_1$] be the precision with which
the value $q$ is known ($\ep(\Q)$ is, say, the mean error of $q$), therefore 
here the wavelength of the light. Let $\ep(\P)$  [originally, $p_1$] be the precision 
with which the value $p$ is determinable; that is, here, the discontinuous change of $p$ 
in the Compton effect \cite[p.~64]{Hei27E}. 
\end{quote}

Heisenberg claimed that this relation is a 
``straightforward mathematical consequence'' \cite[p.~65]{Hei27E} of fundamental 
postulates for quantum mechanics.
In his mathematical derivation of relation \eq{Hei27E}, he derived 
\begin{equation}\label{eq:Hei27S}
\sd{\Q} \,\sd{\P} =\frac{\hbar}{2}
\end{equation}
for standard deviations $\sd{\Q}$ and $\sd{\P}$ of position $\q$ and momentum $\p$
for a class of Gaussian wave functions, later known as minimum uncertainty 
wave packets.
Subsequently, Kennard \cite{Ken27} proved the inequality
\beql{Ken27}
\sd{\Q}\,\sd{\P}\geq\frac{\hbar}{2}
\eeq
for arbitrary wave functions.
By this relation, the lower bound of relation \eq{Hei27E} was later set as
\beql{Hei27K}
\ep(\Q)\ep(\P)\geq\frac{\hbar}{2},
\eeq
where $\hbar=h/(2\pi)$.

Text books \cite{vN55,Boh51,Mes59a,Sch68} up to the 1960s often explained 
that the physical meaning of Heisenberg's uncertainty principle is expressed by 
\Eq{Hei27K},  but it is formally expressed by \Eq{Ken27}.
This explanation is later considered to be confusing.
In fact,  it was said that \Eq{Hei27K} expresses a limitation of measurements,
while mathematically derived relation \Eq{Ken27} expresses 
a statistical property of quantum state,
or a limitation of state preparations, so that they have different meanings \cite{Bal70}.
Thus, Heisenberg with the subsequent completion by Kennard has long been credited 
only with a constraint for state preparation represented by \Eq{Ken27}.

This paper aims to resolve this long standing confusion.
It will be shown that Heisenberg  \cite{Hei27E} in 1927 actually ``proved'' 
not only \Eq{Hei27S} but also \Eq{Hei27E} from basic postulates for quantum mechanics.
In showing that, it is pointed out that as one of the basic postulates Heisenberg supposed 
an assumption called the ``repeatability hypothesis'', which is now considered to be obsolete.
In fact, in the 1930's the repeatability hypothesis was explicitly claimed by von Neumann 
\cite{vN55} and Schr\"{o}dinger \cite{Sch35}, whereas this hypothesis was abandoned in the 1980s, 
when quantum measurement theory was establish to be general enough 
to treat all the physically realizable measurements.

Through those examinations it will be concluded that Heisenberg's uncertainty principle
expressed by \Eq{Hei27K} is logically a straightforward consequence
of \Eq{Ken27} under a generalized form of the repeatability hypothesis.
In fact, under the repeatability hypothesis a measurement is required to prepare 
the state with a sharp value of the measured observable, and hence the  ``measuremental'' 
uncertainty relation \eq{Hei27K} is a logical consequence of the ``preparational''
uncertainty relation \eq{Ken27}.

As stated above, the repeatability hypothesis was abandoned in the 1980s,
and nowadays relation \eq{Hei27K} is taken to be a breakable limit \cite{BK92,GLM04}.
Naturally, the problem remains:
{\em what is the unbreakable constraint for simultaneous measurements of 
non-commuting observables?}
To answer this question, 
we will survey the author's recent proposal \cite{03UVR,03UPQ,04URN}
for a universally valid reformulation of Heisenberg's uncertainty principle 
under the most general assumption on quantum measurement.

\section{Repeatability hypothesis}

The uncertainty principle was introduced by Heisenberg in a paper entitled
{\em \"{U}ber den anschaulichen {Inhalt} der quantentheoretischen {Kinematik}
und {Mechanik}}  \cite{Hei27E} published in 1927.
In what follows we shall examine Heisenberg's derivation of the uncertainty 
principle following this paper.

Before examining the detail of Heisenberg's derivation,
we shall examine the basic postulates for quantum mechanics in Heisenberg's time,
following von Neumann's formulation \cite{vN55}.
In what follows, a positive operator on  a Hilbert space with unit trace is called 
a {\em density operator}.
We denote by $\cB(\R)$ the set of Borel subsets of $\R$
and by $E^{A}$ the spectral measure of a self-adjoint operator $A$, \ie, $A$
has the spectral decomposition $A=\int_\R\la E^{A}(d\la)$.

\begin{Axiom}[\bf States and observables]
Every quantum system $\bS$ is described by a Hilbert space 
${\mathcal H}$ called the {\em state space} of $\bS$. 
{\em States} of $\bS$ are represented by density
operators on $\cH$ and {\em observables} 
of $\bS$ are represented by self-adjoint operators on $\cH$.
\end{Axiom}

\begin{Axiom}[\bf Born statistical formula]
\label{ax:BSF}
If an observable $A$ is measured in a state $\rho$, 
the outcome obeys the probability distribution
of $A$ in $\rho$ defined by
\beqa\label{eq:Born}
\Pr\{A\in\De\|\rho\}=\Tr[E^{A}(\De)\rho],
\eeqa
where $\De\in\cB(\R)$.
 \end{Axiom}

\begin{Axiom}[\bf Time evolution]
Suppose that a system $\bS$ is an isolated system with the (time-independent)
Hamiltonian 
$H$ from time $t$ to $t+\ta$.
The system $\bS$ is in a state  $\rh(t)$ at time $t$ if and only if $\bS$ is in the
state $\rh(t+\ta)$ at time $t+\ta$ satisfying
\beqa
\rh(t+\ta)=e^{-i\ta H/\hbar}\rh(t)e^{i\ta H/\hbar}.
\eeqa
\end{Axiom}

Under the above axioms, we can make a probabilistic prediction of the result of 
a future measurement from the knowledge about the past state.
However, such a prediction applies only to a single measurement in the future.
If we make many measurements successively, we need another axiom
to determine the state after each measurement.
For this purpose, the following axiom was broadly accepted in the 1930s.

\begin{Axiom}[\bf Measurement axiom]
If an observable $A$ is measured in a system $\bS$ to obtain the outcome $a$, 
then the system $\bS$ is left in an eigenstate of $A$ belonging to $a$.
\end{Axiom}

Von Neuamann \cite{vN55} showed that this assumption is equivalent to the following
assumption called the {\em repeatability hypothesis} \cite[p.~335]{vN55}, 
posed with a clear operational condition generalizing a feature of the Compton-Simons 
experiment \cite[pp.~212--214]{vN55}. 
\bigskip

{\bf (R) Repeatability hypothesis.}
{\em If an observable $A$ is measured twice in succession 
in a system $\bS$, then we get the same value each time.}
\bigskip

It can be seen from the following definition of measurement due to Schr\"{o}dinger
given in his famous ``cat paradox'' paper \cite{Sch35} that 
von Neumann's repeatability hypothesis was broadly accepted in the 1930s.

\begin{quote}
The systematically arranged interaction of two systems (measured object and 
measuring instrument) is called a measurement on the first system, if a directly-sensible variable
feature of the second (pointer position) is always reproduced within certain error limits when the
process is immediately repeated (on the same object, which in the meantime must not be exposed 
to any additional influences) \cite{Sch35}.
\end{quote}

Based on the repeatability hypothesis von Neumann \cite{vN55} proved the 
impossibility of simultaneous measurement of two non-commuting observables
as follows.
Suppose that two observables $A, B$ are simultaneously measurable in every
state and suppose that the eigenvalues of $A$ are non-degenerate.    
Then,  the state just after the simultaneous measurement of $A$ and $B$ 
is a common eigenstate of $A$ and $B$, so that
there is an orthonormal basis consisting of common eigenstates of $A$ and $B$,
concluding that $A$ and $B$ commute.

Since Heisenberg's uncertainty principle concerns measurements with errors,
it is naturally expected that it can be mathematically derived by extending the above 
argument to approximate measurements.

\section{Approximate repeatability hypothesis}
\label{se:ARH}

To extend the repeatability hypothesis to approximate measurements,
we generalize the notion of eigenstates as follows.
For any real number $\la$ and a positive number $\ep$,
a (vector) state $\psi$ is called an {\em $\ep$-approximate eigenstate} belonging to $\la$ iff
the relation
\beq
\|A\psi-\la \psi\|\le\ep
\eeq
holds.  If $\ep=0$, the notion of $\ep$-approximate eigenstates is reduced 
to the ordinary notion of eigenstates.  
A real number $\la$ is called an {\em approximate eigenvalue} of an observable $A$
iff for every $\ep>0$ there exists an $\ep$-approximate eigenstate of $A$.
The set of approximate eigenvalues of an observable $A$ coincides with the
spectrum of $A$ \cite[p.~52]{Hal51}.

Now, we formulate the approximate repeatability hypothesis as follows.
\bigskip

{\bf (AR) Approximate Repeatability Hypothesis.}
{\em If an observable $A$  is measured in a system $\bS$ with mean error $\ep$ 
to obtain the outcome $a$, then the system $\bS$ is left in an $\ep$-approximate 
eigenstate of $A$ belonging to $a$.}
\bigskip

Obviously,  (AR) is reduced to (R) for $\ep=0$.
Since we have 
$$
\|A\ps-\la\ps\|\ge\|A\ps-\av{A}\ps\|=\sd{A}
$$ 
for any real number $\la$, where $\av{A}=(\ps,A\ps)$,
(AR) implies the following statement:
{\em If an observable $A$ in a system $\bS$ is measured with mean error $\ep(A)$, 
then the post-measurement standard deviation $\sd{A}$ of $A$ satisfies
\beql{AR}
\sd{A}\le\ep(A).
\eeq}

\section{Heisenberg's derivation of the uncertainty principle}
\label{se:HDUP}

Heisenberg's derivation of \eq{Hei27E} starts with considering a state $\psi$ just after
the measurement of the position observable $\hq$ to obtain the outcome $q'$
with mean error $\ep(\hq)$ and consider what relation holds between $\ep(\hq)$ 
and $\ep(\hp)$ if the momentum observable 
$\hp$ has been measured simultaneously to obtain the outcome $p'$ with mean error $\ep(\hp)$.
Then, by (AR) or \Eq{AR} the state $\psi$ should have the position standard deviation 
$\sd{\hq}$ satsifying 
\beql{SDME}
\sd{\hq}\le \ep(\hq).
\eeq
Heisenberg actually supposed that the state  $\psi$ is a Gaussian wave function \cite[p.~69]{Hei27E}
\beql{MUS}
\psi(q)
=\frac{1}{(\pi q_1^{2})^{1/4}}\exp
\left[-\frac{(q-q')^2}{2q_1^2}-\frac{i}{\hbar}p'(q-q')\right],
\eeq
which is later known as a minimum uncertainty wave packet, 
with its Fourier transform
\beql{FMUS}
\hat{\psi}(p)
=\frac{1}{(\pi p_1^{2})^{1/4}}\exp
\left[-\frac{(p-p')^2}{2p_1^2}+\frac{i}{\hbar}q'(p-p')\right],
\eeq
and he proved relation \eq{Hei27S} for the state $\psi$ given by \Eq{MUS}.

Exactly this part of Heisenberg's argument was generalized by Kennard \cite{Ken27}
to prove relation \eq{Ken27} for any vector state $\psi$.
Thus, Kennard \cite{Ken27} relaxed Heisenberg's assumption on the state $\psi$
to the assumption that 
{\em the state $\psi$ after the position measurement can be arbitrary wave function $\psi$
satisfying \Eq{SDME}.}
Then, if the momentum observable $\hp$ has been measured simultaneously 
to obtain the outcome $p'$ with an error $\ep(\hp)$, by (AR) or \Eq{AR} again 
the state $\psi$ should also satisfy the relation
\beq
\sd{\hp}\le \ep(\hp).
\eeq
Therefore, Heisenberg's uncertainty relation \eq{Hei27K}
immediately follows from Kennard's relation \eq{Ken27}.

As above Heisenberg in 1927 not only derived relation \eq{Hei27E}
by the $\gamma$-ray thought experiment but also gave its mathematical 
proof.   However,  he supposed the repeatability hypothesis or
its approximate version as an additional but obsolete assumption
in addition to the standard postulates for quantum mechanics.

The approximate repeatability hypothesis (AR) has not been explicitly formulated
in the literature, but in the following explanation on the derivation of 
the uncertainty principle von Neumann  \cite[pp.~238--239]{vN55} assumed (AR):
\begin{quote}
We are then to show that if $Q, P$ are two canonically conjugate quantities, 
and a system is in a state in which the value of $Q$ can be given with the
accuracy $\ve$ (\ie, by a $Q$ measurement with an error
range $\ve$), then $P$ can be known with no greater accuracy
than $\et=\hbar/(2\ve)$. Or: a measurement of $Q$ 
with the accuracy $\ve$ must bring about an indeterminacy 
$\et=\hbar/(2\ve)$ in the value of $P$.
\end{quote}
In the above, it is obviously assumed that a state with the position standard deviation 
$\ep$ is resulted by a $Q$ measurement with an error range $\ep$. 
This assumption is what we have generally formulated in \Eq{AR} as 
an immediate logical consequence of  (AR).

Two inequalities \eq{Ken27} and \eq{Hei27K} are often distinguished as
the {\em preparational} uncertainty relation 
and the {\em measuremental} uncertainty relation,
respectively.  However, under the repeatability hypothesis such a distinction 
is not apparent, since a measurement is required to prepare the state 
with a sharp value of the measured observable.  In fact, the above argument
shows that there exists an immediate logical relationship 
between those two inequalities.

\section{Abandoning the Repeatability Hypothesis}
\label{se:ABRH}

The repeatability hypothesis explains only a restricted class of measurements 
and does not generally characterize the state changes caused 
by quantum measurements.
In fact, there exist commonly used measurements of discrete observables,
such as photon counting, that do not satisfy the repeatability hypothesis \cite{IUO90}.
Moreover,  it has been shown that the repeatability hypothesis cannot be generalized 
to continuous observables in the standard formulation of quantum mechanics
\cite{84QC,85CA,Sri80,88MR}.
In 1970, Davies and Lewis \cite{DL70} proposed abandoning the repeatability
hypothesis and introduced a new mathematical framework to treat 
all the physically realizable quantum measurements:
\begin{quote}
One of the crucial notions is that of repeatability which we show is implicitly
assumed in most of the axiomatic treatments of quantum mechanics, but whose
abandonment leads to a much more flexible approach to measurement theory
\cite [p.~239]{DL70}.
\end{quote}

Denote by $\tc(\cH)$ the space of trace class operators on $\cH$,
by $\cS(\cH)$ the space of density operators on $\cH$,
and by $P(\tc(\cH))$ the space of positive linear maps on $\tc(\cH)$.
Davies and Lewis \cite{DL70} introduced a mathematical notion of instrument as follows.
A {\em Davies-Lewis (DL) instrument} for 
(a system $\bS$ described by) a Hilbert space $\cH$ is defined as 
a $P(\tc(\cH))$-valued Borel measure $\cI$ on $\R$ countably additive
in the strong operator topology such that
$\cI(\R)$ is trace-preserving ($\Tr[\cI(\R)\rh]=\Tr[\rh]$).

Let $\bA(\bx)$ be a measuring apparatus for $\bS$ with 
the output variable $\bx$.
The statistical properties of the apparatus $\bA(\bx)$ 
are determined by (i) the probability distribution $\Pr\{\bx\in\De\|\rh\}$
of the outcome $\bx$ in an arbitrary state $\rh$, and
(ii) the state change $\rh\to\rh_{\{\bx\in\De\}}$ from the state $\rh$ 
just before the measurement to the state $\rh_{\{\bx\in\De\}}$ just after the
measurement given the condition $\bx\in\De$.
The proposal of Davies and Lewis \cite{DL70} can be stated as follows.
\bigskip

{\bf  (DL) The Davies-Lewis thesis.}
{\em For every measuring apparatus $\bA(\bx)$ with output variable $\bx$ 
there exists  a unique DL instrument
$\cI$ satisfying
\beqa
\Pr\{\bx\in\De\|\rh\}&=&\Tr[\cI(\De)\rh],
\label{eq:DL1}\\
\rh\to\rh_{\{\bx\in\De\}}&=&
\frac{\cI(\De)\rh}{\Tr[\cI(\De)\rh]}.
\label{eq:DL2}
\eeqa}
\bigskip

For any $\De\in\cB(\R)$, define $\Pi(\De)$ by
\beql{POVM}
\Pi(\De)=\cI(\De)^*1,
\eeq 
where $\cI(\De)^*$ is the dual map of $\cI(\De)$ given 
by $\Tr[(\cI(\De)^{*}X)\rho]=\Tr[X(\cI(\De)\rho)]$
for all $X\in\cL(\cH)$. Then, the map $\De\to \Pi(\De)$ is a probability
operator-valued measure (POVM) \cite{Hel76}, called the {\em POVM of $\cI$},
satisfying
\beql{GBSF}
\Pr\{\bx\in\De\|\rh\}=\Tr[\Pi(\De)\rh]
\eeq
for all $\rh\in\cS(\cH)$ and $\De\in\cB(\R)$.

The problem of mathematically characterizing all the physically 
realizable quantum measurements is reduced to the problem as
to which instruments are physically realizable \cite{04URN}. 
To settle this problem, standard models of
measuring processes were introduced in \cite{84QC} as follows.
A {\em measuring process} for 
(a system described by) a Hilbert space $\cH$
is defined as a quadruple $(\cK,\rh_0,U,M)$ consisting of a Hilbert space
$\cK$, a density operator $\rh_0$ on $\cK$, a unitary operator $U$ on
$\cH\otimes\cK$, and a self-adjoint operator  $M$ on $\cK$.
A measuring process $(\cK,\rh_0,U,M)$ is said to be {\it pure}
if $\rh_0$ is a pure state, and it is said to be {\it separable} if
$\cK$ is separable.

The measuring process $(\cK,\rh_0,U,M)$ mathematically 
models the following description of a measurement.
The measurement is carried out by the interaction, referred to
as the {\it measuring interaction}, between
the {\it object} $\bS$ and the {\it probe} $\bP$.
The probe $\bP$ is described by the Hilbert space $\cK$
and prepared in the state $\rh_0$ just before the measurement.
The time evolution of the composite system $\bP+\bS$ 
during the measuring interaction is described by the unitary operator $U$.
The outcome of the measurement is obtained by measuring 
the observable $M$ called the {\it meter observable} of the probe $\bP$ 
just after the measuring interaction.
We assume that the measuring interaction turns on at time $t=0$ 
and turns off at time $t=\De t$.
In the Heisenberg picture, we write
$$
A_1(0)=A_1\otimes 1, \quad A_2(0)=1\otimes A_2, 
\quad A_{12}(\Det)=U^{\da}A_{12}(0)U,
$$
for an observable $A_1$ of $\bS$, an observable $A_2$ of $\bP$, 
and an observable $A_{12}(0)$ of $\bS+\bP$.

Suppose that the measurement carried out by an apparatus 
$\bA(\bx)$ is described by a measuring process
$(\cK,\rh_0,U,M)$.
Then, it is shown \cite{84QC} that 
the statistical properties of the apparatus $\bA(\bx)$ is given by 
\beqa
\Pr\{\bx\in\De\|\rh\}
&=&\Tr[E^{M(\De t)}(\De)(\rh\otimes\rh_0)],\quad\\
\rh\to\rh_{\{\bx\in\De\}}
&=&\frac{\Tr_{\cK}[(1\otimes E^{M}(\De))U(\rh\otimes\rh_0)U^{\da}]}
{\Tr[E^{M(\De t)}(\De)(\rh\otimes\rh_0)]},
\eeqa
where $\Tr_{\cK}$ stands for the partial trace on the Hilbert space
$\cK$.
The {\em POVM $\Pi$ of the apparatus $\bA(\bx)$} is defined by
\beql{POVM}
\Pi(\De)=\Tr_{\cK}[E^{M(\Det)}(\De)(1\otimes\rh_0)]
\eeq 
for any $\De\in\cB(\R)$.
Then, the map $\De\to \Pi(\De)$ is a probability
operator-valued measure (POVM) \cite{Hel76}
satisfying
\beql{GBSF}
\Pr\{\bx\in\De\|\rh\}=\Tr[\Pi(\De)\rh]
\eeq
for all $\rh\in\cS(\cH)$ and $\De\in\cB(\R)$.

Now it is easy to see that the above description of the measurement statistics 
of the apparatus $\bA(\bx)$ is consistent with the Davies-Lewis thesis.
In fact, the relation 
\beqa\label{eq:instrument_MP}
\cI(\De)\rh=\Tr_{\cK}\left[\left(1\otimes E^{M}(\De)\right)
U(\rh\otimes\rh_0)U^{\da}\right]
\eeqa
defines a DL instrument $\cI$.
In this case, we say that the instrument $\cI$ is {\em realized by the measuring process 
$(\cK,\rh_0,U,M)$.}

A DL instrument for $\cH$ is said to be {\it completely positive (CP)}
if $\cI(\De)$ is completely positive for every $\De\in\cB(\R)$, \ie, 
$\cI(\De)\otimes\id_{n}:
\tc(\cH)\otimes M_{n}\to \tc(\cH)\otimes M_{n}$
is a positive map for every finite number $n$, where $M_n$
is the matrix algebra of order $n$ and $\id_{n}$ is the identity map on $M_n$.
The following theorem characterizes the physically realizable DL instruments 
by completely positivity \cite{83CR,84QC}.

\begin{Theorem}[\bf Realization theorem for CP instruments]
A DL instrument can be realized by a measuring process if and only if
it is completely positive.  In particular, every CP instrument can be
realized by a pure measuring process, and if $\cH$ is separable, 
every CP instrument for $\cH$ can be realized by a pure and separable 
measuring process.
\end{Theorem}

Now, we have reached the following general measurement axiom, abandoning 
Axiom 4 or the repeatability hypothesis.

\begin{Axiom}[\bf General measurement axiom]
\label{ax:GMA}
To every measuring apparatus $\bA(\bx)$ with output variable $\bx$ there exists a unique
CP instrument $\cI$ satisfying Eqs.~\eq{DL1} and \eq{DL2}.
Conversely, to every instrument $\cI$ there exists at least one 
measuring apparatus $\bA(\bx)$  satisfying Eqs.~\eq{DL1} and \eq{DL2}.
\end{Axiom}

\section{Von Neumann's model of position measurement}
\label{se:VNM}

Let $A$ and $B$ be observables of a system $\bS$ described by a Hilbert space $\cH$.
Let $\bA(\bx)$ be a measuring apparatus for $\bS$ with the output variable $\bx$
described by a measuring process $\bM=(\cK,\rho_0,U,M)$ from time $t=0$ to 
$t=\Det$.   An approximate simultaneous measurement of $A(0)$ and $B(0)$ is 
obtained by direct simultaneous measurement of commuting observables 
$M(\Det)$ and $B(\Det)$, where $M(\Det)$ is considered to approximately
measure $A(0)$ and $B(\Det)$ is considered to approximately measure $B(0)$.
In this case the error of the $B(0)$ measurement is called the disturbance of $B$
caused by the measuring process $\bM$, and the relation for the errors of the 
$A(0)$ measurement and the $B(0)$ measurement is called the 
{\em error-disturbance relation (EDR)}.
In what follows, we examine the EDR for position measurement error 
and momentum disturbance.

Until 1980's only solvable model of position measurement had been given 
by von Neumann \cite{vN55}.
We show that this long-standing model satisfies 
Heisenberg's error-disturbance relation \cite{03UVR}, 
a version of Heisenberg's uncertainty relation \eq{Hei27K}.

Consider a one-dimensional mass $\bS$, called an {\em object},
with position $\x$ and momentum $\px$, described by a Hilbert space $\cH=L^{2}(\R_x)$,
where $\R_x$ is a copy of the real line.
The object is coupled from time $t=0$ to $t=\Det$ with the probe $\bP$, 
another one-dimensional mass with position $\y$ and momentum $\py$,  
described by a Hilbert space $\cK=L^2(\R_y)$,  
where $\R_y$ is another copy of the real line.
The outcome of the measurement is obtained by measuring the probe position $\y$ 
at time $t=\Det$.
The total Hamiltonian for the object and the probe is taken to be
\beql{(1)}
{H}_{\bS+\bP} = {H}_{\bS} + {H}_{\bP} + K{H},
\eeq
where ${H}_{\bS}$ and ${H}_{\bP}$ are the free Hamiltonians
of $\bS$ and $\bP$, respectively, ${H}$ represents the measuring interaction.
The coupling constant $K$ satisfies $K\De t=1$ and it is so strong $(K \gg 1)$ 
that ${H}_{\bS}$ and ${H}_{\bP}$ can be neglected.

The measuring interaction $H$ is given by 
\beql{829o} 
H=\x\otimes\py,
\eeq 
so that the unitary operator of the time evolution of $\bS+\bP$ from $t=0$ to $t=\ta\le \Det$ is given by
\beql{829p}
U(\ta)=\exp\left(\frac{-iK\ta}{\hbar}\x\otimes\py\right).
\eeq

Suppose that the object $\bS$ and the probe $\bP$ are in the vector states $\psi$ and $\xi$, 
respectively, just before the measurement.
We assume that the wave functions $\ps(x)$ 
and $\xi(y)$ are Schwartz rapidly decreasing functions \cite{RS80}.
Then, the time evolution of $\bS+\bP$ in the time interval  $(0,\De t)$ 
is given by the unitary operator
$
U(\De t)=e^{-i\hat{x}\otimes\hat{p}_{y}/\hbar}.
$
Thus, this measuring process is represented by
$(L^{2}(\R_y),\ketbra{\xi},e^{-i\hat{x}\otimes\hat{p}_{y}/\hbar},\hat{y})$.

The state of the composite system $\bS+\bP$ just after the measurement is 
$U(\Det){\ps\otimes\xi}$.
By solving the Schr\"{o}dinger equation,  we have
\beq
U(\Det)(\ps\otimes\xi)(x,y)=\ps(x)\xi(y-x).
\eeq
From this, 
the probability distribution of output variable $\bx$ is given by
\beqa
\Pr\{\bx\in\De\|\psi\}=\int_{\De}\, dy\int_{\R} |\psi(x)|^2\,|\xi(y-x)|^2\, dx.
\eeqa
By a property of convolution, if the probe probability distribution $|\xi(y)|^2$ approaches 
to the Dirac delta function $\de(y)$, 
the output probability approaches to the Born probability distribution $|\psi(x)|^2$.

The corresponding instrument $\cI$ is given by
\beql{INS-VNM}
\cI(\De)\rh=\int_{\De}\xi(y1-\hat{x})\rho\xi(y1-\hat{x})^{\da}dy,
\eeq
and the corresponding POVM is given by 
\beql{POVM-VNM}
\Pi(\De)=\int_{\De}|\xi(y1-\hat{x})|^{2}dy,
\eeq

Solving the Heisenberg equations of motion, 
we have
\beqa
\x(\De t)&=&\x(0),\label{eq:S1}\\
\y(\De t)&=&\x(0)+\y(0),\label{eq:S2}\\
\px(\De t)&=&\px(0)-\py(0),\label{eq:S3}\\
\py(\De t)&=&\py(0).\label{eq:S4}
\eeqa      

\section{Root-mean-square error and disturbance}
\label{se:RMS}

To define the ``mean error'' of the above position measurement,
let us recall classical definitions.
Suppose that a quantity $X=x$ is measured by 
directly observing another quantity $Y=y$.   
For each pair of values $(X,Y)=(x,y)$, 
the error is defined as $y-x$.  
To define the ``mean error'' given 
the joint probability distribution (JPD)
$ \mu^{X,Y}(dx,dy)$ of $X$ and $Y$, 
Gauss \cite{Gau95} introduced the {\em root-mean-square (rms) error} 
$\ep_{G}(X,Y)$ of $Y$ for $X$ as
\beql{rmse}
\ep_{G}(X,Y)=\left(\iint_{\R^{2}}
(y-x)^{2} \mu^{X,Y}(dx,dy)\right)^{1/2},
\eeq
which Gauss  \cite{Gau95} called the ``mean error'' or the ``mean error to be feared'',
and has long been accepted as a standard definition for  the ``mean error''. 

In the von Neumann model, the observable $\x(0)$ is measured 
by directly observing the meter observable $\y(\De t)$.
Since $\x(0)$ and $\y(\De t)$ commute by \Eq{S2},
we have the JPD 
$\mu^{\x(0),\y(\De t)}(dx,dy)$ of $\x(0)$ and $\y(\De t)$
as
\beql{JPDE}
\mu^{\x(0),\y(\De t)}(dx,dy)=\bracket{E^{\x(0)}(dx)E^{\y(\De t)}(dy)},
\eeq
where 
$\bracket{\cdots}$ stands for the mean value in the state $\ps\otimes\xi$.
Then,  by \Eq{rmse} the {\em rms error} $\ep(\x,\psi)$  
for measuring $\x$ in state $\psi$ is defined as the rms error $\ep_G(\x(0),\y(\De t))$
of $\y(\De t)$ for $\x(0)$, so that we have 
\beqa
\ep(\x,\psi)
&=&\left(\iint_{\R^{2}}(y-x)^{2}\mu^{\x(0),\y(\De t)}(dx,dy)\right)^{1/2}\nn\\
&=&\bracket{(\y(\De t)-\x(0))^{2}}^{1/2}
=\bracket{\y(0)^{2}}^{1/2}.
\label{eq:E-V}
\eeqa

Since $\px(0)$ and $\px(\De t)$ also commute from \Eq{S3},
we also have the JPD
$\mu^{\px(0),\px(\De t)}(dx,dy)$ of the values of $\px(0)$ and $\px(\De t)$.
The {\em rms disturbance} $\et(\px,\psi)$ of $\px$ 
in state $\psi$ is similarly defined as the rms error $\ep_G(\px(0),\px(\De))$,
so that we have
\beqa
\et(\px,\psi)
&=&\left(\iint_{\R^{2}}(y-x)^{2}\mu^{\px(0),\px(\De t)}(dx,dy)\right)^{1/2}\nn\\
&=&\bracket{(\px(\De t)-\px(0))^{2}}^{1/2}
=\bracket{\py(0)^2}^{1/2}.
\label{eq:D-V}
\eeqa

Then, by Kennard's inequality \eq{Ken27} we have
\beqa
\ep(\x,\psi)\et(\px,\psi)&=&\bracket{\y(0)^{2}}^{1/2}\bracket{\py(0)^2}^{1/2}\nn\\
&\ge& \si(\y(0))\si(\py(0))\ge\frac{\hbar}{2}.
\eeqa
Thus, the von Neumann model satisfies Heisenberg's error-disturbance relation
(EDR) 
\beql{HEDR-QP}
\ep(\x)\et(\px)\ge\frac{\hbar}{2}
\eeq
for $\ep(\x)=\ep(\x,\psi)$ and $\et(\px)=\et(\px,\psi)$.

By the limited availability for measurement models up to the 1980's,
the above result appears to have enforced a prevailing belief 
in Heisenberg's EDR \eq{HEDR-QP}, for instance,  
in claiming the standard quantum limit for gravitational wave detection 
\cite{BVT80,CTDSZ80,Cav85}.

\section{Measurement Violating Heisenberg's EDR}
\label{se:MVH}

In 1980,  Braginsky,  Vorontsov, and Thorne \cite{BVT80} claimed that
Heisenberg's EDR \eq{HEDR-QP} leads  to 
a sensitivity limit, called the {\em standard quantum limit} (SQL), 
for gravitational wave detectors exploiting free-mass position monitoring.
Subsequently,  Yuen \cite{Yue83} questioned the validity of the SQL,
and then Caves \cite{Cav85} defended the SQL by giving 
a new formulation and a new proof without directly appealing 
to Heisenberg's ERD \eq{HEDR-QP}.
Eventually,  the conflict was reconciled \cite{88MS,89RS} by pointing out that 
Caves \cite{Cav85} still supposed (AR),
in spite of avoiding Heisenberg's ERD \eq{HEDR-QP}.
More decisively, a solvable model of a precise position measurement was also
constructed that breaks the SQL \cite{88MS,89RS}; later  
this model was shown to break Heisenberg's EDR \eq{HEDR-QP} \cite{02KB5E}.

In what follows, we examine this model, which 
modifies the measuring interaction of the von 
Neumann model. 
In this new model, the object, the probe, and the probe
observables, the coupling constant  $K$, and the time duration  $\De t$
are the same as the von Neumann model. 
The measuring interaction is taken to be \cite{88MS}
\beql{829ox}
H=\frac{\pi}{3\sqrt{3}}
(2\x\otimes\py-2\px\otimes\y
+\x\px\otimes 1-1\otimes \y\py).
\eeq
The corresponding instrument is give by \cite{04URN}
\beq
\cI(\De)\rh=\int_{\De}e^{-ix\px}\ketbra{\phi}e^{-ix\px}\Tr[E^{\x}(dx)\rh],
\eeq
where $\phi(x)=\xi(-x)$, and the corresponding POVM is given by
\beq
\Pi(\De)=E^{A}(\De).
\eeq

 Solving the Heisenberg equations of motion, 
 we have
\beqa
\x(\De t)&=&\x(0)-\y(0),\label{eq:ozawa-model-1}\\
\y(\De t)&=&\x(0),\label{eq:ozawa-model-2}\\
\px(\De t)&=&-\py(0),\label{eq:ozawa-model-3}\\
\py(\De t)
&=&\px(0)+\py(0)\label{eq:ozawa-model-4}.
\eeqa

Thus,  $\x(0)$ and $\y(\De t)$ commute and also $\px(0)$ and $\px(\De t)$ commute,
so that the rms error and the rms disturbance are well defined by their JPDs, and given by
\beqa
\ep(\x,\psi)
&=&0,\\
\et(\px,\psi)
&=&\bracket{(\py(0)+\px(0))^2}^{1/2}<\infty.
\eeqa
Consequently,  we have 
\beq
\ep(\x)\et(\px)=0.
\eeq
Therefore, this model obviously violates Heisenberg's EDR \eq{HEDR-QP}.  

\section{Universally Valid Error-Disturbance Relation}
\label{se:UVE}

To derive a universally valid EDR, 
consider a measuring process $\bM=(\cK,\rh_0,U,M).$
If $A(0)$ and $M(\Det)$ commute, 
the rms error of the measuring process $\bM$ 
for measuring $A$ in $\rh$ can be defined  
through the JPD of $A(0)$ and $M(\Det)$.  
Similarly, if $B(0)$ and $B(\Det)$ commute, the rms disturbance 
can also be defined through the JPD of $B(0)$ and $B(\Det)$.  
In order to extend the definitions of the rms error and disturbance
to the general case, we introduce the noise operator and the disturbance operator.

The {\em noise operator} $N(A)$ is defined as 
the difference $M(\Det)- A(0)$ between the observable $A(0)$ to be measured
and the meter observable $M(\Det)$ to be read and 
the {\em disturbance operator} $D(A)$ is defined as the 
the change $B(\Det)-B(0)$ of $B$ caused by the measuring interaction, 
\ie,
\beqa
N(A)&=&M(\Det)- A(0),\\
D(B)&=&B(\Det)-B(0).
\eeqa
The {\em mean noise operator} $n(A)$ and the {\em mean disturbance operator} $d(B)$ 
are defined by
\beqa
n(A)&=&\Tr_{\cK}[N(A)1\otimes \rh_0],\\
d(B)&=&\Tr_{\cK}[D(B)1\otimes \rh_0].
\eeqa
The {\em rms error} $\ep(A,\rh)$ 
and the {\em rms disturbance} $\et(B,\rh)$ 
for observables $A,B$, respectively, in state $\rh$ are defined by
\beqa
\ep(A,\rh)&=&(\Tr[N(A)^2\rho\otimes \rh_0])^{1/2}, \\
\et(B,\rh)&=&(\Tr[D(B)^2\rho\otimes \rh_0])^{1/2}.
\eeqa
An immediate meaning of $\ep(A,\rh)$ and  $\et(B,\rh)$ are the rms's 
of the noise operator and the disturbance operator.

Suppose that $M(\De  t)$ and $A(0)$ commute in $\rho\otimes\rh_0$, \ie,
\beq
[E^{A(0)}(\De),E^{M(\Det)}(\Ga)]\rho\otimes\rh_0=0
\eeq
for all $\De,\Ga\in\cB(\R)$ \cite{Gud68,05PCN,06QPC}.  
In this case, the relation
\beq
\mu^{A(0),M(\Det)}(dx,dy)=\Tr[E^{A(0)}(dx)E^{M(\Det)}(dy)\rh\otimes\rh_0]
\eeq
defines the JPD of $A(0)$ and $M(\Det)$ satisfying 
\beq
\Tr[p(A(0),M(\Det))\rh\otimes\rh_0]=\iint_{\R^{2}}p(x,y)\,\mu^{A(0),M(\Det)}(dx,dy)
\eeq
for any real polynomial $p(A(0),M(\Det))$ in $A(0)$ and $M(\Det)$ \cite{Gud68}.
Thus, the classical rms error $\ep_G(A(0),M(\Det))$ of $M(\Det)$ for $A(0)$ 
is well defined, and we easily obtain the relation
\beq
\ep(A,\rh)=\ep_G(A(0),M(\Det)).
\eeq
Similarly, we have $\et(B,\rh)=\ep_G(B(0),B(\Det))$ if $B(0)$ and $B(\Det)$ commute
in $\rh\otimes\rh_0$.

In 2003, the present author \cite{03HUR,03UVR,03UPQ} derived the  relation
\beql{UEDR}
\ep(A) \et(B)+|\bracket{[n(A),B]}+\bracket{[A,d(B)]}|
\ge  \frac{1}{2}\left| \langle  [A,B]  \rangle \right|,
\eeq
where $\ep(A)=\ep(A,\rh)$, $\et(B)=\et(B,\rh)$,
which is universally valid for any observables $A,B$, any system state $\rh$, 
and any measuring process $\bM$.
From \Eq{UEDR}, it is concluded that if the error and the disturbance are statistically 
independent from system state,  then the Heisenberg type EDR
\beql{HEDR}
\ep(A)\et(B)\geq\frac{1}{2}|\bracket{[A,B]}|
\eeq
holds, extending the previous results \cite{AG88,Ray94,91QU,Ish91}.
The additional correlation term in \Eq{UEDR} allows the error-disturbance product 
$\epsilon (A) \eta (B)$ to violate the Heisenberg type EDR \eq{HEDR}.
In general,  the relation
\beql{OEDR}
\epsilon(A) \eta(B)+\epsilon(A)\sigma(B)+\sigma(A)\eta(B) 
\ge  \frac{1}{2}\left| \langle  [A,B]  \rangle \right|,
\eeq
holds for any observables $A,B$, any system state $\rh$, and 
any measuring process $\bM$
\cite{03HUR,03UVR,03UPQ,04URJ,04URN,05UUP}.

The new relation \eq{OEDR} leads to the following 
new constraints for precise measurements
and non-disturbing measurement: 
then
\beqa\label{eq:DB}
\sigma(A)\eta(B) &\ge&  \frac{1}{2}\left| \langle  [A,B]  \rangle \right|,\quad\mb{if $\ep(A)=0$}, \\
\epsilon(A)\sigma(B)&\ge&  \frac{1}{2}\left| \langle  [A,B]  \rangle \right|,
\quad\mb{if $\et(B)=0$}. \label{eq:EB}
\eeqa
Note that if $\bracket{[A,B]}\not=0$, the 
Heisenberg type EDR \eq{HEDR} leads to the divergence of $\ep(A)$ or $\et(B)$
in those cases.
The new error bound \Eq{EB} was used to derive a conservation-law-induced limits for
measurements \cite{02CLU,03UPQ,04UUP,BL11}
quantitatively generalizing the 
Wigner-Araki-Yanase theorem \cite{Wig52,AY60,Yan61,91CP}
and was used to derive a fundamental accuracy limit for quantum computing \cite{03UPQ}. 

\section{Quantum Root Mean Square Errors}
\label{QRMS}

We say that the measuring process $\bM$ is {\em probability reproducible}  
for the observable $A$ in the state $\rh$ iff 
\beq
\Tr[E^{M(\Det)}(\De)\rho\otimes\rho_0]=\Tr[E^{A}(\De)\rh]
\eeq 
holds for all $\De\in\cB(\R)$.
The rms error $\ep(A,\rh)$ satisfies that
$\rh(A,\rh)=0$ for all $\rh$ if and only if 
$\bM$ is probability reproducible for $A$ in all $\rh$ \cite{02KB5E,04URN}.
Thus, the condition that $\ep(A,\rh)=0$ for all $\rh$ characterizes the class of
measurements with POVM $\Pi$ satisfying $\Pi=E^{A}$.

Busch, Heinonen, and Lahti \cite{BHL04} pointed out that there are cases
where $\ep(A,\rh)=0$ holds but $\bM$ is not probability reproducible 
and where $\bM$ is not probability reproducible but $\ep(A,\rh)=0$ holds,
and questioned the reliability of the rms error $\ep(A.\rh)$ as a state-dependent error measure.
However, their argument lacks a reasonable definition of precise measurements,
necessary for discussing the reliability of error measures.
In response to their criticism, the present author \cite{05PCN,06QPC}
has successfully characterized the precise measurements of $A$
in a given state $\rh$ and shown that the rms error $\ep(A,\rh)$ reliably 
characterizes such measurements. 
In what follows we survey those results, which were mostly neglected 
in the recent debates \cite{DN14,KJR14,BLW14RMP}.

Let us start with the classical case.
Suppose that a quantity $X=x$  is measured by direct observation of  
another quantity $Y=y$.  
Then, this measurement is {\em precise} iff  $X=Y$
holds with probability 1, or equivalently the JPD 
$\mu^{X,Y}(dx,dy)$ of $X$ and $Y$ concentrates on 
the diagonal set, \ie, 
\beq
\mu^{X,Y}(\{(x,y)\in\R^2\mid x\ne y\})=0.
\eeq
As easily seen from \Eq{rmse},
this condition is equivalent to the condition $\ep_G(X,Y)=0$. 

Generalizing the classical case, we say that a measuring process $\bM$ makes a 
{\em strongly precise measurement} of $A$ in $\rh$ 
iff $A(0)=M(\Det)$ holds with probability 1
in the sense that $A(0)$ and $M(\Det)$ commute in $\rh\otimes\rh_0$ and that 
the JPD $\mu^{A(0),M(\Det)}$ concentrates on the diagonal set,
\ie, 
\beq
\mu^{A(0),M(\Det)}(\{(x,y)\in\R^2\mid x\ne y\})=0.
\eeq
On the other hand, we have introduced another operational requirement. 
The {\em weak joint distribution} $\mu_{W}^{A(0),M(\De t)}$ 
of $A(0)$ and $M(\De t)$ in a state $\rh$ is defined  by
\beq
\mu_{W}^{A(0),M(\De t)}(dx,dy)=\Tr[E^{A(0)}(dx)E^{M(\De t)}(dy)\rho\otimes\rho_0].
\eeq
The weak joint distribution is not necessarily positive 
but operationally accessible by weak measurement
and post-selection \cite{LW10}.
We say that the measuring process $\bM$ makes a {\em weakly precise measurement}
of $A$ in $\rh$ iff
the weak joint distribution $\mu_{W}^{A(0),M(\De t)}$ in state $\rh$ 
concentrates on the diagonal set, \ie,
\beq
\mu_{W}^{A(0),M(\De t)}(\{(x,y)\in\R^2\mid x\ne y\})=0.
\eeq
This condition does not require that $A(0)$ and $M(\Det)$ commute, while it only requires
that the weak joint distribution concentrates on the event $A(0)=M(\Det)$.
A similar condition has been used to observe momentum transfer in a double-slit 
`which-way' experiment \cite{GWPP04,MLMSGW07}. 
We naturally consider that strongly preciseness is a sufficient condition for
precise measurements and weak preciseness is a necessary condition.
In the previous investigations \cite{05PCN,06QPC},  it was mathematically 
proved that both conditions are equivalent.  Thus, either condition is concluded 
to be a necessary and sufficient condition characterizing the unique class 
of precise measurements.  
As above, we say that the measuring process $\bM$ {\em precisely measures}
$A$ in $\rh$ iff it makes a strongly or weakly precise measurement of $A$ in $\rh$.

To characterize the class of precise measurements in terms of 
the {\em rms error-freeness condition}, $\ep(A,\rh)=0$,
and the probability reproducibility condition, we introduce the following notions.
The {\em cyclic subspace} $\cC(A,\rh)$ generated by $A$ and $\rh$ is defined  
as the closed subspace of $\cH$
generated by $\{E^{A}(\De)\phi\mid \De\in\cB(\R), \phi\in\ran (\rh)\}$, 
where 
$\ran(\rh)$ denotes the range of $\rh$.
Then, the following theorem holds \cite{05PCN,06QPC}.

\begin{Theorem}
Let $\bM=(\cK,\rh_0,U,M)$ be a measuring process 
for the system $\bS$ described by a Hilbert space $\cH$.
Let $A$ be an observable of $\bS$ and $\rh$ a state of $\bS$. 
Then, the following conditions are equivalent.

{\rm (i)} $\bM$ precisely measures $A$ in $\rh$.

{\rm (ii)}  $\ep(A,\phi)=0$ in all $\phi\in\cC(A,\rh)$.

{\rm (iii)} $\bM$ is probability reproducible for  $A$ in all $\phi\in\cC(A,\rho)$. 
\end{Theorem} 

In the case where $A(0)$ and $M(\Det)$ commute, precise measurements are characterized by
the rms error-freeness condition, since in this case we have
$\ep_G(A(0),M(\Det))=\ep(A,\rh)$.
However, the probability reproducible condition does not characterize
the precise measurements even in this case.
To see this suppose that $A(0)$ and $M(\Det)$ are identically distributed and independent 
\cite[p.~763]{06QPC}.
Then, we have
\beqas
\lefteqn{
\ep_G(A(0),M(\Det))=\iint_{\R^{2}}(y-x)^{2} \mu^{A(0)}(dx) \mu^{M(\Det)}(dy)}\\
&=&\si(A(0))^2+\si(M(\Det))^2+(\av{A(0)}-\av{M(\Det)})^2.
\eeqas
Since $\si(A(0))=\si(M(\Det))$ and $\av{A(0)}=\av{M(\Det)}$, we have
\beq
\ep_G(A(0),M(\Det))=\sqrt{2}\si(A).
\eeq
Thus, $\bM$ is not a precise measurement for the input state $\rh$ with $\si(A)\ne 0$.
In the case where $A(0)$ and $M(\Det)$ do not commute, the rms error-freeness
condition well characterizes precise measurements to a similar extent 
to the probability reproducibility condition. 
In particular,  the class of measuring processes precisely measuring $A$ in all $\rh$
is characterized by the following equivalent conditions \cite{05PCN,06QPC}: 
(i)  $\ep(A,\psi)=0$ for all $\psi\in\cH$;
(ii) probability reproducible for  $A$ in all $\psi\in\cH$; 
(iii) $\Pi=E^{A}$.
The above result ensures our long-standing belief that a measurement with POVM $\Pi$ 
satisfying $\Pi=E^{A}$ is considered to be precise in any state in the sense that the measured observable $A(0)$ 
and the meter observable $M(\Det)$ to be directly observed are perfectly 
correlated in any input state, not only reproducing the probability distribution in any state.

We say that the measuring process $\bM$ {\em does not disturb} an observable $B$ in a state $\rh$ 
iff observables $B(0)$ and $B(\De t)$ commute in the state $\rh\otimes\rh_0$ 
and the JPD $\mu^{B(0),B(\De t)}$ of  $B(0)$ and $B(\De t)$ concentrates on the diagonal set.
The non-disturbing measuring processes defined above can be characterized analogously.

From the above results, a non-zero lower bound for $\ep(A)$ or $\et(B)$ indicates impossibility of precise or 
non-disturbing measurement.
In particular, if $\si(A),\si(B)<\infty$ and $\bracket{[A,B]}\ne 0$, then 
any measuring process cannot precisely measure $A$ without disturbing $B$.

The above characterizations of precise and non-disturbing measurements 
suggests the following definitions 
of the {\em locally uniform rms error} $\overline{\ep}(A,\rh)$
and the {\em locally uniform rms disturbance} $\overline{\et}(B,\rh)$
\cite{06NDQ}: 
\beqa
\overline{\ep}(A,\rh)&=&\sup_{\phi\in\cC(A,\rh)}\ep(A,\phi),\\
\overline{\et}(B,\rh)&=&\sup_{\phi\in\cC(B,\rh)}\et(B,\phi).
\eeqa
Then, we have
$\overline{\ep}(A,\rh)=0$ if and only if the measurement precisely
measures $A$ in $\rh$, and that $\overline{\et}(B,\rh)=0$ if and only if the measurement does not disturb $B$ in $\rh$.  
For those quantities, the Heisenberg type EDR 
\beql{EDR-LU}
\overline{\ep}(\x)\overline{\et}(\px)\ge\frac{\hbar}{2}
\eeq
is still violated by a linear position measurement \cite{06NDQ}, and 
the relation
\beqa
\overline{\ep}(A)\overline{\et}(B)+\overline{\ep}(A)\sigma(B)
+\sigma(A)\overline{\et}(B)\geq\frac{1}{2}|
	\bracket{[A,B]}|
\eeqa
holds universally \cite{06NDQ}, where
$\overline{\ep}(A)=\overline{\ep}(A,\rh)$ and
$\overline{\et}(B)=\overline{\et}(B,\rh)$.

Thus, the locally uniform rms error $\overline{\ep}(A,\rh)$
completely characterizes precise measurements of $A$ in $\rh$
and the locally uniform rms disturbance $\overline{\et}(B,\rh)$
completely characterizes measurements non-disturbing $B$ in $\rh$,
while they satisfy the EDR of the same form as the rms error and 
disturbance.  
Further investigations on quantum generalizations of the classical
notion of root-mean-square error and EDRs formulated with those
quantities will be reported elsewhere.

\begin{acknowledgements}
This work was supported in part by JSPS KAKENHI, No.~26247016
and No.~15K13456, and the John Templeton Foundation, ID \#35771.
\end{acknowledgements}

\end{document}